\documentclass{optica-article}
\journal{opticajournal} 

\articletype{Research Article}

\usepackage{lineno}
\usepackage{eurosym}
\begin{document}

\title{Cost-effective time-stretch terahertz recorders using 1550 nm probes}

\author{Christelle Hanoun\authormark{1}, Eléonore Roussel\authormark{1}, Christophe Szwaj\authormark{1}, Clément Evain\authormark{1}, Marc Le Parquier\authormark{1,2}, Jean-Blaise Brubach\authormark{3}, Nicolas Hubert\authormark{3}, Marie Labat\authormark{3}, Pascale Roy\authormark{3}, Marie-Agnès Tordeux, and Serge Bielawski\authormark{1}\authormark{*}}

\address{\authormark{1}Univ. Lille, CNRS, UMR 8523 - PhLAM - Physique des Lasers Atomes et Mol\'ecules, F-59000 Lille, France\\
\authormark{2}Centre d'Etudes et de Recherches Lasers et Applications (CERLA), F-59000 Lille, France\\
\authormark{3}Synchrotron SOLEIL, Synchrotron SOLEIL, Gif-sur-Yvette, France}
\email{\authormark{*}serge.bielawski@univ-lille.fr} 


\begin{abstract*} 
Time-stretch electro-optic detection allows THz waveforms to be recorded in single-shot, up to Megahertz acquisition rates. This capability is required in accelerator physics, and  also opens new applications in table-top THz time-domain spectroscopy. However, the technique has also been notoriously known for the need of high speed -- and high cost -- ADCs or oscilloscopes for the readout. Furthermore, the resulting cost considerably increases with the number of samples that is needed per THz waveform, an issue that has severely limited the widespread of the technique so far. In this article, we show that particularly cost-effective designs can be obtained by using 1550~nm probes. We present the performances of an experimental design, that uses only standard components (including dispersive devices), and a standard commercial probe laser without additional pulse broadening. In these conditions, our time-stretch system could already record coherent THz pulses at the SOLEIL synchrotron radiation facility, over an unprecedented number of samples, using oscilloscopes and ADC boards with only 1-3 GHz bandwidth.
\end{abstract*}

\section{Introduction}
There is an increasing demand for methods allowing ultrafast optical signals to be recorded in single shot, and at multi-Megahertz acquisition rates, using femtosecond probes. Applications include the analysis of pulsed sources in the optical~\cite{hammer2016single,herink2017real}, X-ray~\cite{diez2023sensitive} and THz~\cite{roussel2015observing} domains. Such methods are also required in a range of cutting-edge spectroscopic applications, where samples under interest present fast and non-reproducible dynamics~\cite{kobayashi2019fast,Couture2023,couture2023performance}, and in order to speed up analyses in two-dimensional spectroscopy~\cite{gao2023ultrafast,donaldson2023breaking}. Single-shot methods are also key in research and operation of accelerators, synchrotron light sources, and free-electron lasers that operate at multi-MHz repetition rates~\cite{evain2017direct,steffen2020compact,funkner2019high,funkner2023revealing,ilyakov2022field}.

For achieving such single-shot recordings in the THz domain, at high repetition rates (i.e., millions of traces per second) an increasingly popular technique is photonic time-stretch electro-optic detection~\cite{roussel2015observing,kobayashi2019fast,Couture2023,couture2023performance}, 
a technique that is directly inspired by the so-called photonic time-stretch Analog-to-Digital Converter~\cite{bhushan1998time}. The electric signal of interest modulates a chirped laser pulse, using the electro-optic effect in a Pockels crystal. Then, the modulated laser pulse is further stretched until it can be recorded by a conventional photodetector and a fast oscilloscope or ADC board. As a result, the recorded optical pulse is a replica of the electric field under investigation, that is temporally stretched by a factor $M$, known as the (temporal) stretch factor. Compared to other single-shot THz recording techniques~\cite{Kawada:11,Shan:00,Minami.2013} time-stretch THz digitizers are -- by design -- able to operate at high repetition rates.

However, achieving THz photonic time-stretch typically requires high bandwidth -- and therefore high cost -- oscilloscopes for the readout. This high cost has therefore limited applications of the technique to research in large instruments~\cite{roussel2015observing,bielawski2019self}, and pioneer demonstrations in time-domain spectroscopy~\cite{kobayashi2019fast,Couture2023}. For reducing the cost of THz recording systems to reasonable values, a key point is to reduce further the bandwidth required for the readout. Several studies have been performed for overcoming this bottleneck, in particular by performing time-stretch using long chirped Bragg gratings~\cite{kobayashi2019fast}. Another solution consists in avoiding time-stretch by using spectrally-decoded electro-optic detection based on ultrafast linear detector arrays~\cite{rota2019kalypso,steffen2020compact,funkner2023revealing}. Although these methods have been demonstrated to be efficient at MHz repetition rates and above, the limited availability of key technological components (several meter-long Bragg gratings and multi-MHz rate cameras) are still limiting their use for the moment.

In this Letter, we present a design that allows Terahertz photonic time-stretch electro-optic detection to be performed with much longer stretch durations, for a given laser bandwidth, by using standard commercial components. A key point of the strategy consists of avoiding the classical wavelengths for Terahertz Electro-Optic Detection -- mainly 800 nm and 1030 nm -- and operating at the 1550~nm wavelength. Before going further, it is important to note that, surprisingly, few studies of single-shot 1550~nm-based THz electro-optic detection have been performed so far~\cite{roussel2023single}. This contrasts with the fact that the 1550~nm wavelength has been widely used for many other applications of the so-called \emph{time-stretch ADC} since its introduction in 1998~\cite{bhushan1998time}. We will show here that -- in addition to be a viable option -- the association of 1550~nm-based THz electro-optic detection~\cite{nagai2004generation,schneider2006high,roussel2023single,wilke2024thin} with time-stretch presents major advantages with respect to previous high repetition rate electro-optic detection systems. As a first consequence, much lower costs will immediately be obtained, by enabling the use of few~GHz oscilloscopes for the readout. Furthermore, we will see that even 1~GHz bandwidth readout will provide larger effective number of samples to be recorded, compared to similar systems operating at the more popular 1030~nm wavelength. Tests will be performed by recording THz pulses emitted at the AILES THz beamline of the SOLEIL synchrotron radiation facility, at a repetition rate of 0.85~MHz.

\section{Experimental setup}
\subsection{Time-stretch THz recorder design}
\begin{figure}[htbp]
\begin{center}
\includegraphics[width=\textwidth]{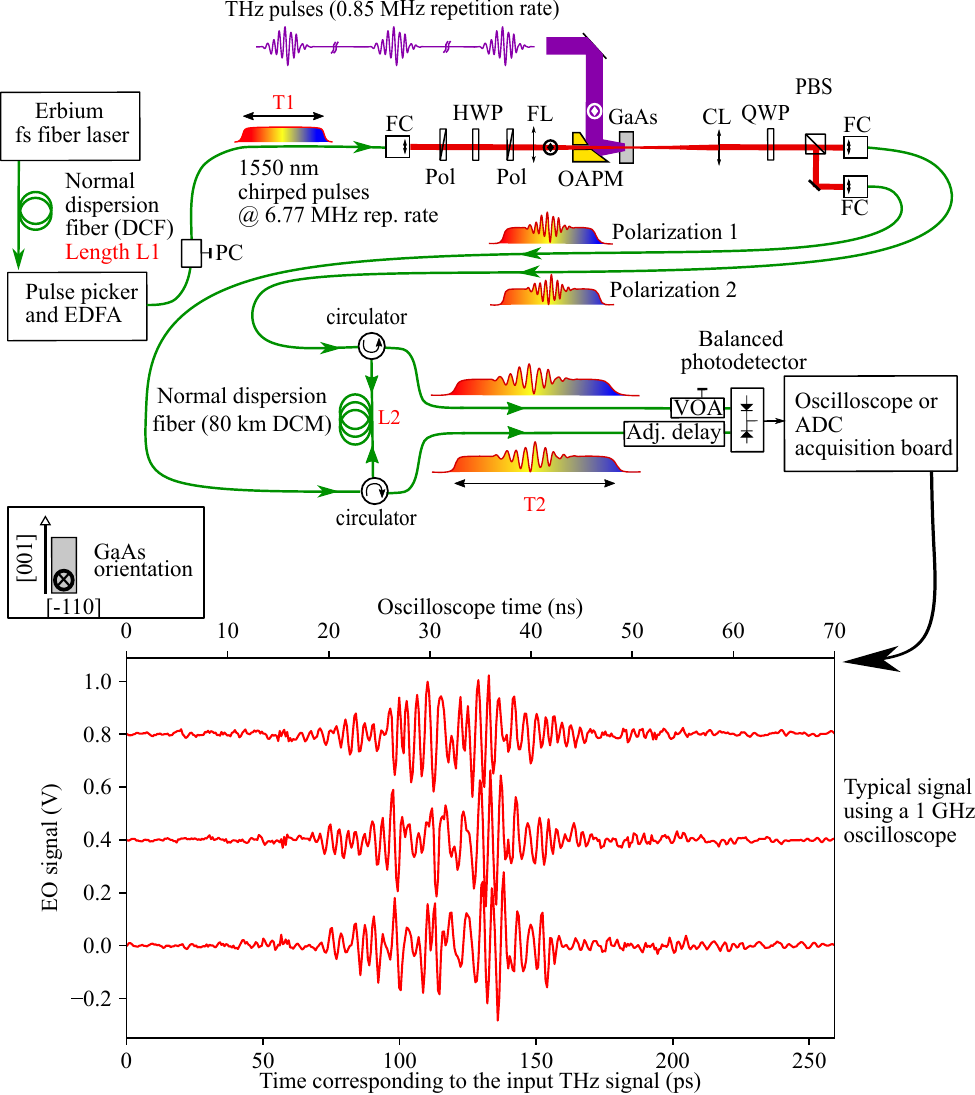}
\end{center}
\caption{Time-stretch setup and typical THz signals recorded in single-shot. THz pulses are focused on a Gallium Arsenide (GaAs) Pockels crystal, and interact with 1550~nm chirped probe laser pulses. After the polarizing beam-splitter (PBS), the two laser pulses are intensity-modulated. The second stretch is achieved by a normal dispersion fiber-based telecommunication Dispersion-Compensation Module (DCM, with a dispersion equivalent to 80~km of standard single-mode fiber). Green: single-mode fiber (standard telecommunication fiber, except the DCF and DCM). EDFA: SM Erbium-doped fiber amplifier, PC: polarization controller, FC: fiber collimator with 6.18~mm focal length, OAPM: 50~mm off-axis parabolic mirror, HWP and QWP: achromatic half-wave and quarter-wave plates, FL and CL: identical lenses with 100~mm focal length used for focusing and collimation,  VOA: variable optical attenuator. The QWP is adjusted at 45$^\circ$ with respect to the THz and laser polarizations. The displayed electro-optic (EO) signals (vertically shifted for clarity) have been recorded at the AILES beamline of Synchrotron SOLEIL.}
\label{fig:exp_setup}
\end{figure}

The experimental setup is displayed in Figure~\ref{fig:exp_setup}. Chirped probe pulses are delivered by a femtosecond Erbium fiber laser (Menlo C-Fiber, with 100~mW output power, 68~nm bandwidth, and 88.05~MHz repetition rate). The laser pulses are first chirped using 40~m of normal dispersion fiber (-140 ps/nm/km dispersion-compensation fiber). Then, the pulse repetition rate is divided by 13 using an acousto-optic pulse picker (Gooch \&~Housego Fiber-Q). After amplification, the resulting probe pulses have 6.77~MHz repetition rate, and typically 100~mW average power and $30$~nm FWHM bandwidth. The laser repetition rate is synchronized to the main RF clock of the SOLEIL facility, using an RRE synchronization module from Menlo, thus ensuring the synchronization of the probe pulses (at 6.77~MHz) with respect to the delivered THz pulses (at 0.85~MHz repetition rate). Note that, in this test, the acquisition rate of the detection system ($6.77\times 10^6$ measurements per second) was overdimensioned with respect to the 0.85~MHz repetition rate of the THz source.

Each THz pulse (at 0.85 MHz repetition rate) modulates a probe pulse (one every eight), using the Pockels effect in a 3~mm-thick, [110]-cut Gallium Arsenide crystal, and a relatively standard polarization optics setup~\cite{roussel2015observing}. The two differential outputs are stretched by the same amount, by using a single mode fiber and a set of two circulators (with <0.8~dB loss). The final stretch is provided by low-cost commercial telecommunication dispersion compensation module, based on a DCF fiber (DCM-80, from Fiberstore), with 1360~ps/nm dispersion, and $\le 5.8$~dB~loss. We perform a balanced detection of the two stretched outputs, using a 2.5~GHz InGaAs balanced photoreceiver (Thorlabs PDB780CAC, with $5000$~V/A differential gain and 15~pW/$\sqrt{\text{Hz}}$ noise-equivalent input power). For recording the data, we used two devices. A 1~GHz oscilloscope (Lecroy WaveSurfer 4104HD) has been used for the development and preliminary tests of the setups. Then, systematic data recordings have been performed at the THz beamline of Synchrotron SOLEIL, using essentially a 3~GHz bandwidth ADC board (Teledyne SP Devices ADQ7DC, PCIe version with 10~GS/s sampling rate and 4~GB memory). The price ranges of both the 1~GHz oscilloscope and the 3~GHz ADC board were similar, and well below the cost of the probe laser.

\subsection{THz coherent synchrotron source}
SOLEIL is a synchrotron radiation facility, that is based on a 2.75~GeV electron storage ring with 354~m circumference. One or several relativistic electron bunches can be stored in the storage ring (one in the present case), and emit light at various wavelengths (from the THz to the X-ray domains). Most of the emission is usually incoherent. However, an intense coherent emission (Coherent Synchrotron Radiation, or CSR) can also be produced, if a density modulation can be applied in the longitudinal direction of an electron bunch. Remarkably, such a modulation (and coherent emission) can occur naturally, as a result of a self-organization process, known as the microbunching instability~\cite{byrd2002observation,abo2003brilliant}. This instability simply requires the bunch charge to exceed a threshold value. These two last decades, there has been an important activity aiming at understanding the underlying mechanisms of the coherent emission, including theoretical modeling, direct observations~\cite{roussel2015observing,rota2019kalypso,bielawski2019self,funkner2023revealing}, and control~\cite{evain2019stable,wang2021accelerated,evain2023stabilization}. The coherent THz emission is constituted of pulses at the repetition frequency of the electron bunch (0.85~MHz), that is modulated by a slower envelope (typically at few kHz frequency). This coherent radiation is delivered at the AILES beamline~\cite{barros2013coherent}.
For the present test, the parameters of the storage ring are close to the situation of references~\cite{roussel2015observing,szwaj2016high}. The similarity of conditions enables to compare the capabilities of the present 1550~nm probe laser-based time-stretch measurement system, with respect to previous systems working at the 1030~nm wavelength.

\section{Results}
\label{sec:results}
\begin{figure}
    \centering
    \includegraphics[width=\textwidth]{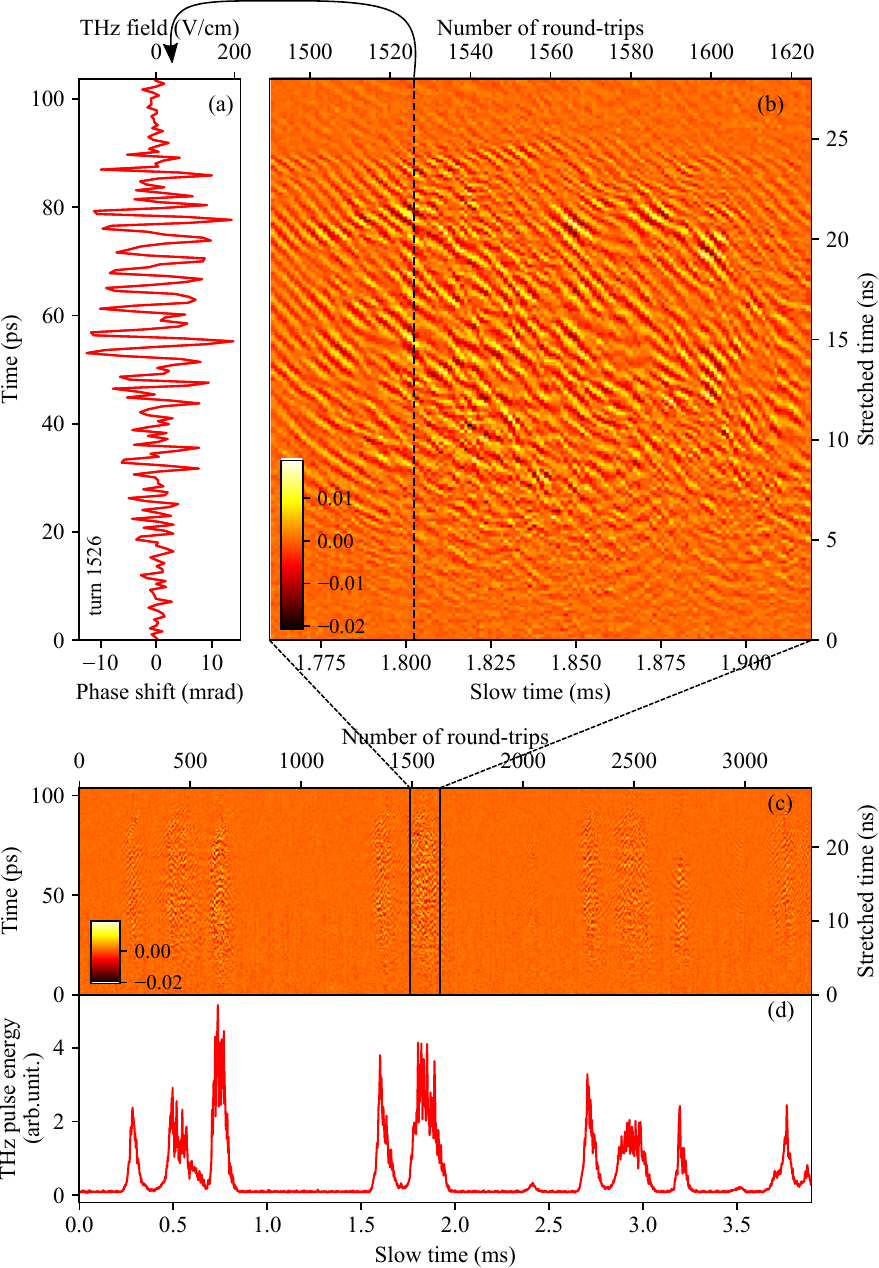}
    \caption{Series of giant coherent THz pulses recorded in one shot at SOLEIL, during a study of the so-called microbunching instability. (a) Electro-optic signal of a single THz pulse. (b) Series of successive THz pulses displaying characteristic details of the dynamics during a THz burst (including carrier-envelope phase-drift and spatio-temporal dislocations). (c) Same data plotted over a longer time period, displaying the characteristic bursting behavior of the THz emission. (d): corresponding THz pulse energy versus pulse number. The electro-optic signal is expressed as the THz-induced phase-shift in the electro-optic crystal (see~Eq.\ref{eq:E_vs_phase_shit}). Note that the acquisition rate of the time-stretch recording system (6.77~MHz) is overdimensioned with respect to the repetition rate of the source (0.85~MHz). One record contains approximately 170000 THz pulses (only part of the data has been displayed for clarity). These data have been recorded using a 3~GHz ADC acquisitition board (see text).}
\label{fig:signals}
\end{figure}

\begin{figure}
\includegraphics[width=\textwidth]{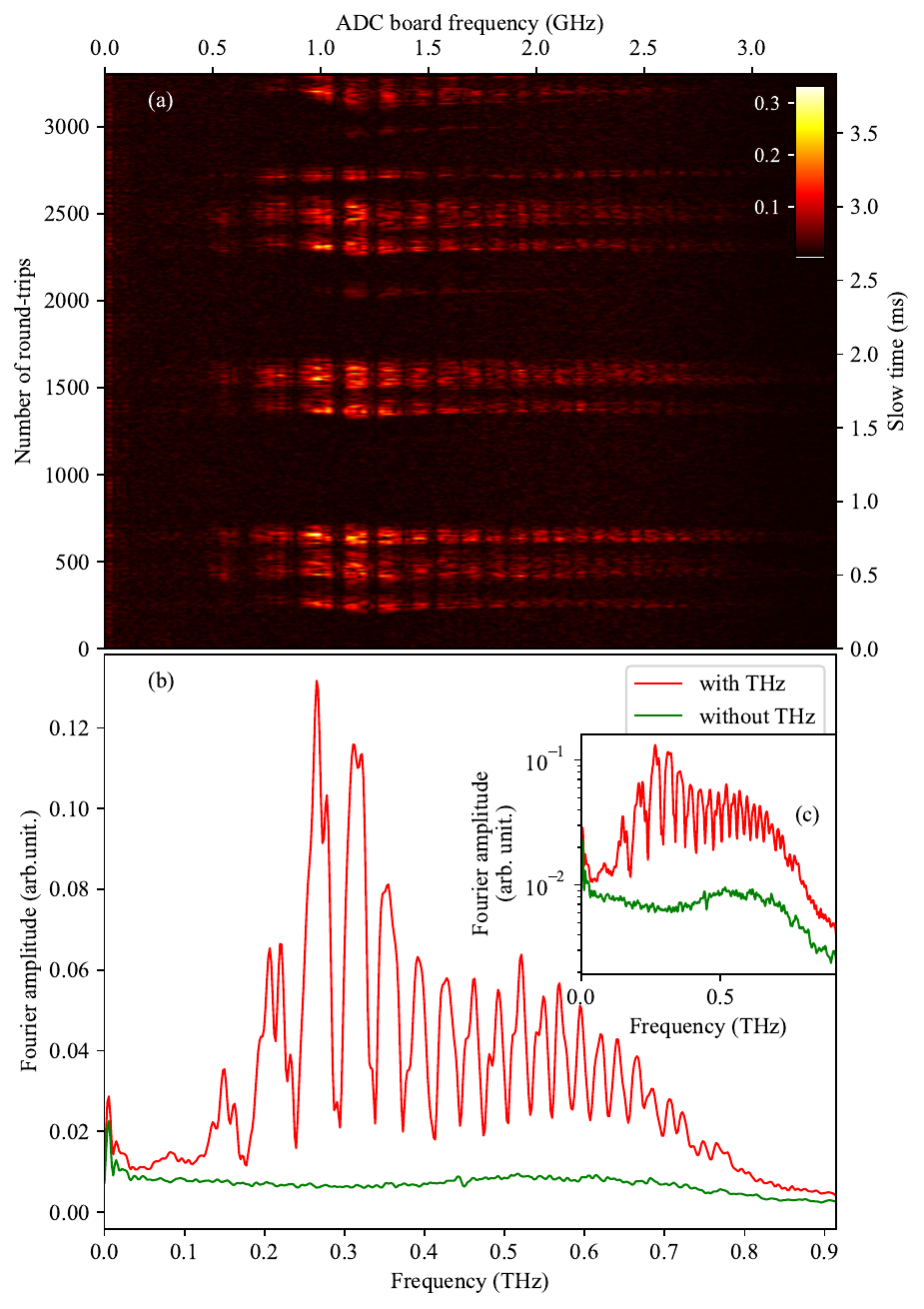}
\caption{Fourier spectra of the THz pulses displayed in Figure~\ref{fig:signals}. (a) THz amplitude spectra represented as a colorscale diagram versus pulse number and THz frequency (each spectrum is obtained from a single pulse). The time-evolution corresponds to the bursting behavior expected from the physics of the Coherent Synchrotron Radiation emission (the microbunching instability, see text). Note that the spectra present a series of dips, that are due to the dispersion penalty phenomenon associated to the measurement technique (see Ref.~\cite{deos}).  (b) Average amplitude spectrum over all round-trips (c) Inset: Average Fourier spectrum in logarithmic scale. Green line: detection noise level (average amplitude spectrum without input THz signal).}
\label{fig:FT_signals}
\end{figure}

A typical series of THz pulses is represented in Figure~\ref{fig:signals}, and the corresponding THz spectra are represented in Figure~\ref{fig:FT_signals}. As expected from the photonic time-stretch process, the THz signals appear on the oscilloscope, as "stretched versions" of the input THz signal, by a stretch factor of $M=270$ here. These data have been obtained using a classical processing. 
The balanced photodetector signal is first re-sampled (i.e., interpolated) at a multiple of the laser repetition rate. Then we extract the balanced electro-optic signal~$\Delta V_n(t)$ corresponding to THz pulse shot $n$, and subtract the background (the average over 3 unmodulated balanced detection signals before and after the THz pulse). An example of signal at this preprocessing stage is displayed in Figure~\ref{fig:exp_setup}.

Then a rather classical processing (mainly a normalization by the average laser pulse shape) provides the THz-induced birefringence phase-shift $\Delta\phi_n(t)$ (see, e.g., Ref.~\cite{szwaj2016high}, Section~3):
\begin{eqnarray}
    \Delta\phi_n(t)=\frac{\Delta V_n(t)}{2V_L(t)},
\end{eqnarray}
where $V_L(t)$ is the average laser pulse shape (recorded separately, see Fig.~\ref{fig:noise}b).

From the phase shift~$\Delta\phi(t)$, we estimate the electric field~$E_{THz}(t)$ inside the crystal using the classical expression~\cite{szwaj2016high}:
\begin{equation}
\Delta\phi(t)=\frac{2\pi d}{\lambda}n_0^3r_{41}E_{THz}(t),\label{eq:E_vs_phase_shit}
\end{equation}
where $d=3$~mm is the crystal thickness, and $\lambda=1550$~nm is the laser wavelength. $n_0=3.38$ and $r_{41}=1.5$~pm/V~\cite{nagai2004GaAs} are the GaAs refractive index (at the laser wavelength) and Pockel coefficient, respectively. A typical evolution of the birefringence phase-shift~$\Delta\phi(t)$ (and deduced electric field~$E_{THz}(t)$) corresponding to a typical THz pulse is represented in Figure~\ref{fig:signals}a. 

Grouping the THz pulses in a colorscale map then gives a representation of the shot-to-shot dynamics (Figures~\ref{fig:signals}b~and~\ref{fig:signals}c)). As in previous observations~\cite{roussel2015observing,bielawski2019self}, the dynamics occurs in bursts of THz emission (see Figures~\ref{fig:signals}c~and~\ref{fig:signals}d)).

Figure~\ref{fig:FT_signals}a displays the single shot THz spectra, corresponding to the EO signals of Figure~\ref{fig:signals}, i.e., over more than 3000 round-trips. The bursting behaviour of the THz emission can be clearly seen in the colorscale map representation. The average spectrum over all round-trips is represented in Fig.~\ref{fig:FT_signals}b. All measured signals reveal a broadband spectrum with a central frequency around $0.3$~THz. The spectra present nulls at specific frequencies that correspond to the dispersion penalty phenomenon, typically observed in photonic time-stretch digitizers~\cite{han2005phasediv}.

It is important to note that, although the obtained data look very similar to  previous work, considerable progress is obtained at two levels. Figure~\ref{fig:signals} shows that we were not only able to reduce the readout ADC cost, but also that we considerably extended the recording duration (i.e., the probe laser duration $T_1$)), which enables to study the full THz CSR pulses at each turn in the storage ring. A more detailed and fair comparison with previous work however requires a more precise conceptual framework, and this will be the subject of the next Section.

\section{Discussion}

\subsection{Discussion with respect to previous achievements}
The results show that a $\approx 35-40$~ns stretch (FWHM) could be obtained, using this detector design, a commercial 1550~nm laser source and a basic erbium aplifier (i.e, without spectral broadening). Before entering details, we can see that this already represents an advantage with respect to previously reported values using 1030~nm commercial oscillators (without spectral broadening, i.e., as in the present experiment): $\approx 4$~ns without amplification~\cite{roussel2015observing}, and $\approx 2$~ns after amplification~\cite{evain2017direct}.

Furthermore, this advantage is likely to evolve in the near future, because the final stretch duration does not only depend on the achievable fiber dispersion, but also on the achievable probe laser bandwidth. Further progress is therefore expected, by adding external pulse broadening, such as parabolic amplification and supercontinuum generation, in the spirit of previous achievements using 1030~nm probes~\cite{steffen2020compact,couture2023performance}.

Given this variety of designs (already existing and foreseen), it becomes important to define reliable numbers, that allows a fair comparison between different THz recorder setups. 

\subsection{Effective number of samples and time-bandwidth product}
\begin{figure}
\begin{center}
\includegraphics[width=\textwidth]{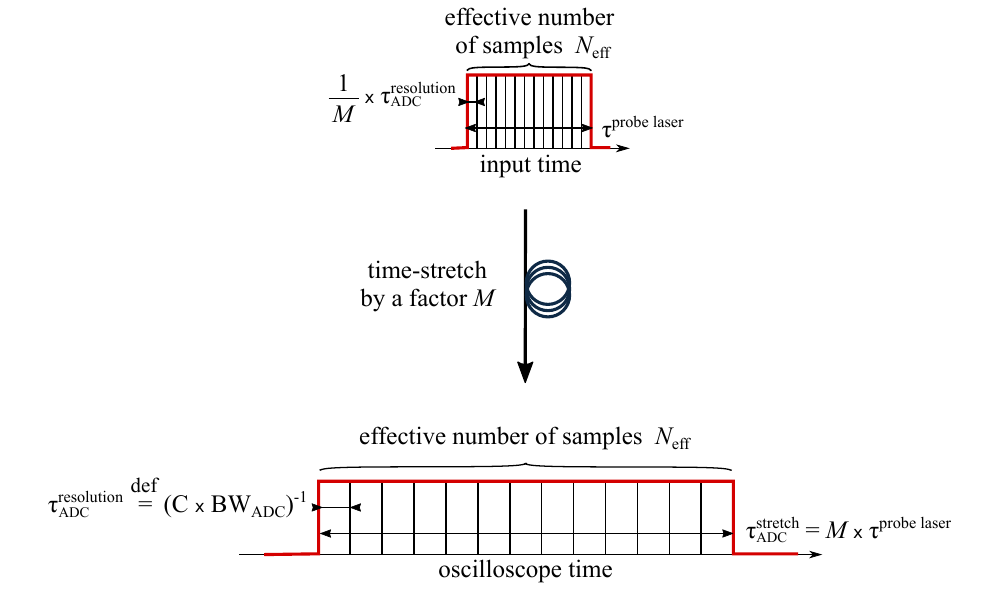}
\end{center}
\caption{Illustration of the quantities involved in the definition in Eqns.~(\ref{eq:Neff_def}-\ref{eq:Neff_Neff1_TBP}). The effective number of samples $N_\text{eff}$ is proportional (by a constant $C$ depending only on the shape of the impulse response of the readout) to the time-bandwidth product: $N_\text{eff,0}=BW_\text{ADC}\times\tau^\text{stretch}_\text{ADC}$. $N_\text{eff,0}$ is thus a convenient quantity for comparing the effective number of points of different time-stretch systems. 
}
\label{fig:illustration_resolution}
\end{figure}

Using a low-bandwidth (for instance few GHz) digitizer  is not, by itself, an argument in favor of a given design. A more fair comparison point is the {\it number of samples} $N_\text{eff}$ that can be recorded, for a given oscilloscope (and front-end electronics) bandwidth budget $BW_\text{ADC}$ (see Figure~\ref{fig:illustration_resolution}):
\begin{eqnarray}
N_\text{eff}=\frac{\tau^\text{stretch}_\text{ADC}}{\tau^\text{impulse}_\text{ADC}},\label{eq:Neff_def}
\end{eqnarray}
where $\tau^\text{impulse}_\text{ADC}$ defines the "time resolution" ($\tau^{\text{resolution}}_\text{ADC}$ in Figure.~\ref{fig:illustration_resolution}) of the electronics (from photodetector to ADC), and $\tau^\text{stretch}_\text{ADC}$ denotes the laser pulse duration at the photodetector. We have
\begin{eqnarray}
\tau^\text{stretch}_\text{ADC}\approx\Delta\lambda D_2L_2,
\end{eqnarray}
where $\Delta\lambda$ is the laser probe bandwidth, and $L_2$ and $D_2$ are the dispersion and length of the second fiber (see Figure~\ref{fig:exp_setup}, where $\tau^\text{stretch}_\text{ADC} \equiv T_2$). Note that this approximation is valid for THz time-stretch recorders, where the stretch factor is typically extremely large, and $\tau^\text{stretch}_\text{ADC}$ is negligibly affected by the length of the $L_1$ fiber.

As in any low-pass electronics, for a fixed shape of the transfer function, the impulse response duration $\tau^\text{impulse}_\text{ADC}$ is inversely proportional to the bandwidth, i.e.:  
\begin{eqnarray}
\tau^\text{impulse}_\text{ADC}=\frac{1}{C\times BW_\text{ADC}},
\end{eqnarray}
where the $C$ parameter depends on the shape of impulse response, and the definition of the bandwidth and rise/fall times (e.g., $C=2\pi$ for the textbook case where the recorder has a first order response, $BW_\text{ADC}$ is the 3dB bandwidth, and $\tau^\text{impulse}_\text{ADC}$ is the $1/e$ pulse width). The effective number of recorded samples can be therefore directly expressed as a time-bandwidth product at the readout electronics:
\begin{eqnarray}
&&N_\text{eff}=C \times N_\text{eff,0}\\
&&\text{with\;}N_\text{eff,0}=\tau^\text{stretch}_\text{ADC}\times BW_\text{ADC}.\label{eq:Neff_Neff1_TBP}
\end{eqnarray}
We propose to use the time-bandwidth product $N_\text{eff,0}$ as a figure-of-merit for comparing different THz time-stretch recorder designs (keeping in mind that the effective number of samples $N_\text{eff}$ will be typically larger that $N_\text{eff,0}$ by few units).

\subsection{Effective number of samples (time-bandwidth product $N_\text{eff,0}$) for several designs}
In table~\ref{tab:stretch}, we compared the time-bandwidth products $N_\text{eff,0}$ for several designs. The calculation is made using laser bandwidths of previous THz time-stretch experiments. For fair comparison, we assumed the same attenuation in the second fiber $L_2$ than in the present experiment (6~dB), and the same recording bandwidth (2.5~GHz, the bandwidth of the balanced photoreceiver).

The present arrangement leads to a value of $N_\text{eff,0}$ approximately 5 times larger than with a standard 1030~nm commercial laser, with 40~nm bandwidth (e.g., the laser of Ref.~\cite{roussel2015observing}), and a stretch in a HI1060 fiber. 

Further improvement of $N_{\text{eff}}$ can be performed using spectral broadening of the laser pulses, using nonlinear propagation in a fiber. This has been achieved in a THz time-stretch detector at 1030~nm, using nonlinear propagation in a fiber, which led to 100~nm bandwidth~\cite{Couture2023}. It is interesting to note that even though we did not use any additional spectral broadening for the moment (and a modest 30~nm bandwidth), the value of $N_{\text{eff}}$ is already two times larger than for the 1030~nm/100~nm bandwidth case.

\begin{table}
\begin{center}
\begin{tabular}{|c|c|c|c|c|c| } 
 \hline
 Probe laser type & Stretch fiber, loss& $L_2$& $\tau^\text{stretch}_\text{ADC}$  & $N_\text{eff,0}$ for  \\
  and bandwidth $\Delta\lambda$&  and dispersion & for 6 dB &   &  $BW_\text{ADC}$    \\
  && loss&  & $=2.5$~GHz \\
\hline
Yb fiber laser& HI1060 & & &\\
 (1030 nm) $\Delta\lambda$=40 nm &1.5~dB/km, -53 ps/nm/km &4 km& 8.5 ns& 21 \\ 
 \hline
Yb laser (1030~nm) with & HI1060& & & \\
NL broadening, $\Delta\lambda$=100 nm  & 1.5~dB/km, -53 ps/nm/km & 4 km& 21 ns&53\\ 
\hline
Erbium laser (1550~nm) & SMF28 & &&\\
  $\Delta\lambda$=30 nm& 0.17 dB/km, 18 ps/nm/km& 35 km&19 ns & 48\\ 
\hline
{\bf Erbium laser, 1550~nm,}& {\bf Dispersion compensation} & {\bf Corresp. to}& &\\
   {\bf  $\Delta\lambda$=30 nm} & {\bf fiber. 6~dB loss, 1350~ps/nm} & {\bf 80 km SMF} & {\bf 40~ns} &{\bf 101}  \\ 
 \hline\end{tabular}
\end{center}
\caption{Comparison of the effective number of recorded points, corresponding to several key designs. For a clear comparison, in the three first cases, we displayed calculated values, assuming the same detection bandwidth (2.5~GHz), and a length of stretch fiber inducing the same loss (6~dB) than in our experiment (last row). The Ytterbium fiber laser bandwidth corresponds to the Menlo Orange laser used in Ref.~\cite{roussel2015observing}. The spectrally broadened Ytterbium laser corresponds to Ref.~\cite{Couture2023}. In bold: present experimental performance (note that the used laser has a 70~nm bandwidth, that is reduced to $\approx 30$~nm bandwidth after amplification). Also note that the actual number of points $N_\text{eff}$ is actually larger than the time-bandwidth product $N_\text{eff,0}$ by few units (see text).}
\label{tab:stretch}
\end{table}

\subsection{Sensitivity of the time-stretch detection system}
\begin{figure}
\begin{center}
\includegraphics[width=\textwidth]{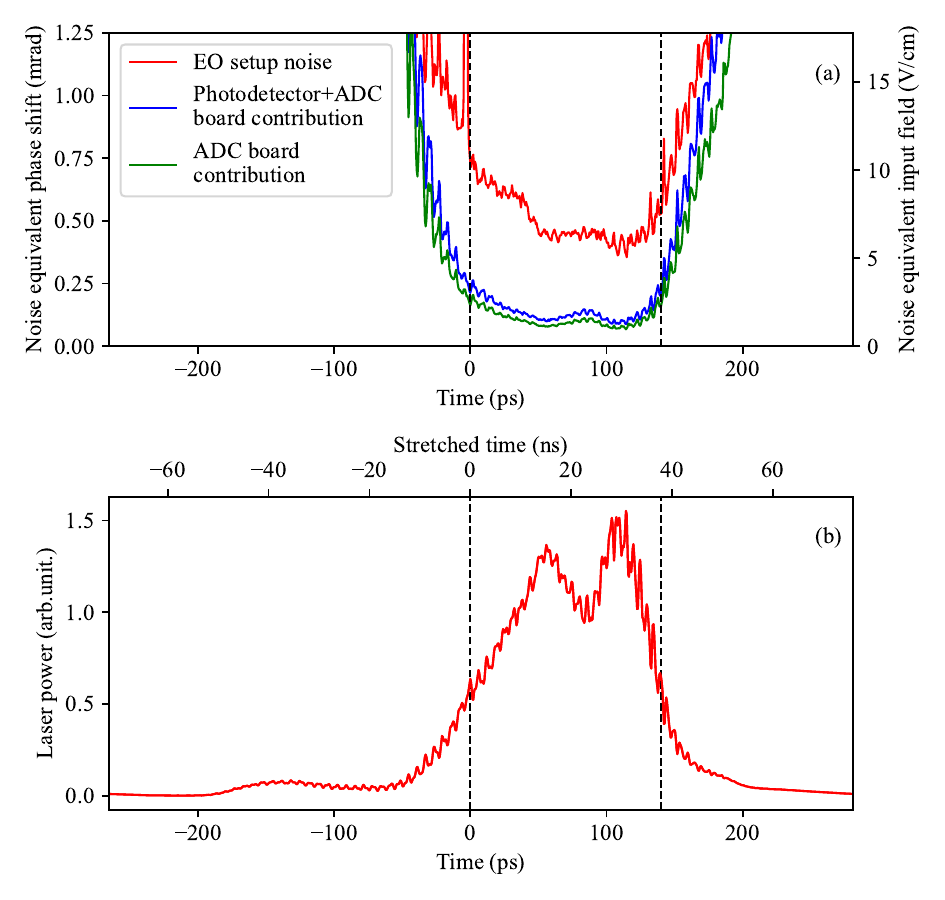}
\end{center}
\caption{Sensitivity of the Electro-Optic (EO) measurement setup. (a): sensitivity, expressed as the noise-equivalent input signal (red line). The blue line is the noise obtained when the laser is OFF (i.e., the noise contributions from the detector and ADC board). The green curve corresponds to the ADC board only. The left vertical axis is the Pockels phase-shift, i.e., corresponding to the THz field-induced birefringence in the GaAs crystal.
The right axis is an estimation of the electric field in the crystal, assuming plane waves and a constant electro-optic coefficient $r_{41}$ vs frequency (see text). (b): laser power versus time. The signals used for Figures~\ref{fig:signals} and \ref{fig:FT_signals} correspond to the time interval between the two vertical dashed lines.}
\label{fig:noise}
\end{figure}

The sensitivity can be defined by the input signal level, that corresponds to the detection background noise (i.e., without input THz signal). The corresponding curves are plotted in Figure~\ref{fig:noise}, together with the noise contributions of the detector and ADC board. Near the probe laser peak, the noise equivalent Pockels phase-shift $\Delta\phi_{noise}$ is typically below 0.5~mrad (RMS). An estimate of the corresponding  electric field $E_{noise}$ can be obtained from Eq.~(\ref{eq:E_vs_phase_shit}) (keeping in mind that this assumes plane waves for the laser and THz signal in a first approximation). The corresponding sensitivity is typically below 7~V/cm inside the crystal.

Note that, although the equivalent input noise may seem much larger 
than in previous work ($\approx 0.1$~mrad or 1.25V/cm in 
Ref.~\cite{szwaj2016high}), it important to note that the two 
situations are very different in terms of effective number of point. $N_\text{eff}\approx 100$ here, whereas $N_\text{eff}\approx 3$ for the 2~ns stretch and 1.5~GHz bandwidth for the noise measured in Ref.~\cite{szwaj2016high} (i.e., over a 300~GHz bandwidth in that context). 

\section{Conclusion}
We presented a new design for time-stretch THz electro-optic detection, that has the potential to widespread the development of time-stretch THz electro-optic detection. As a key point, using a probe at the 1550~nm wavelength (instead, in particular, of the popular 1030~nm wavelength) allows significantly larger stretch factors to be obtained in a simple way, using commercial femtosecond lasers without additional spectral broadening. As a first advantage, a considerable reduction of the overall cost is directly obtained, while keeping a relatively simple design. In the presented application, the oscilloscope and ADC board cost (of the order of 15~k\euro{} as of 2024), is more than one order of magnitude lower than in previous time-stretch THz recorders~\cite{roussel2015observing,szwaj2016high,evain2017direct,Couture2023,couture2023performance}. As a second advantage, for a fixed oscilloscope budget (i.e., oscilloscope bandwidth), the increase in stretch factor directly translates into an increase in the effective number of recorded samples. 

It is also important to note that, in this work, we focused on design strategies leading to longer stretching. However further capabilities will be worth to be  integrated, in particular diversity electro-optic sampling (DEOS) for improving the signal fidelity~\cite{deos}. In addition, it is worth keeping in mind that we used a laser source with moderate optical bandwidth (30~nm after amplification). Hence, the full potential of the technique has yet to be explored, by adding nonlinear pulse broadening techniques to the strategy (in a similar way to what has been done using 1030~nm probes~\cite{Couture2023}).

Finally, it is worth noticing that using 1550~nm probes is foreseen as a potential advantage in accelerator facilities, most of which are already hosting femtosecond timing systems at 1550~nm. 
On the other hand, an open question concerns the feasibility of single-shot time-domain spectrometers using 1550~nm lasers, for both the THz generation and the time-stretch detection, in the spirit of the recent work performed at other wavelengths~\cite{kobayashi2019fast,Couture2023}.

\begin{backmatter}
\bmsection{Funding}

The project has been supported by the Ministry of Higher Education and Research, Nord-Pas
de Calais Regional Council and European Regional Development Fund (ERDF) through
the Contrat de Plan
\'Etat-R\'egion (CPER photonics for society), the LABEX CEMPI project (ANR-11-LABX-0007), and the ANR-DFG ULTRASYNC project (ANR-19-CE30-0031). ER was supported by the METEOR CNRS MOMENTUM grant.


\bmsection{Disclosures}




\noindent The authors declare no conflicts of interest.




\bmsection{Data availability}
The data that support the plots within this paper and
other findings of this study are available from the corresponding
author upon reasonable request.












\end{backmatter}

\end{document}